%%%%%%%%%%%%%%%%%%%%%%% preprint.sty % renomme prepiop.sty %%%%%%%%%
%
%%%%%% Preprint style for articles to be submitted to IOP journals
%
%%%%%%% Prepared by AJC 12/2/88, updated 1/2/89 and by SW 12/2/91
%%%%%%% Last modified by JFM 05/05/92
%%%%%%%%%%%%%%%%%%%%%%%%%%%%%%%%%%%%%%%%%%%%%%%%%%%%%%%%%%%%%%%%%%%%%
%
% First we have a character check
%
% ! exclamation mark    " double quote  
% # hash                ` opening quote (grave)
% & ampersand           ' closing quote (acute)
% $ dollar              % percent       
% ( open parenthesis    ) close paren.  
% - hyphen              = equals sign
% | vertical bar        ~ tilde         
% @ at sign             _ underscore
% { open curly brace    } close curly   
% [ open square         ] close square bracket
% + plus sign           ; semi-colon    
% * asterisk            : colon
% < open angle bracket  > close angle   
% , comma               . full stop
% ? question mark       / forward slash 
% \ backslash           ^ circumflex
%
% ABCDEFGHIJKLMNOPQRSTUVWXYZ 
% abcdefghijklmnopqrstuvwxyz 
% 1234567890
%
%%%%%%%%%%%%%%%%%%%%%%%%%%%%%%%%%%%%%%%%%%%%%%%%%%%%%%%%%%%%%%%%%%%%%

%%%%%%% Dimensions (page width is width of journal page x1.2) %%%%%%%

\magnification=1200
\hsize=31pc
\vsize=55 truepc
\hfuzz=2pt
\vfuzz=4pt
\pretolerance=5000
\tolerance=5000
\parskip=0pt plus 1pt
\parindent=16pt
%

%%%%%%%%%%%%%%%%%%%%%%%%% Font definitions %%%%%%%%%%%%%%%%%%%%%%%%%%
%
% Fonts for title
%
\font\fourteenrm=cmr10 scaled \magstep2
\font\fourteeni=cmmi10 scaled \magstep2
\font\fourteenbf=cmbx10 scaled \magstep2
\font\fourteenit=cmti10 scaled \magstep2
\font\fourteensy=cmsy10 scaled \magstep2

% Font for small caps within title and authors names
%
\font\large=cmbx10 scaled \magstep1

% Font for matrices (please replace with cmbx10 
% if you do not have this font available)
%

% Font for matrices (use \bss{x} for bold sans serif within maths)
%

% Font for vectors (bold italic). Please replace with cmbx10
% if you do not have this font available.
% Use \bi{r} for bold italic r within maths 
%

% Fonts for small type (if used)
%
\font\eightrm=cmr8  
\font\eighti=cmmi8
\font\eightbf=cmbx8
\font\eightit=cmti8

\font\eightsy=cmsy8
\font\sixrm=cmr6
\font\sixi=cmmi6
\font\sixsy=cmsy6

%%%% Definitions of tenpoint, eightpoint and fourteenpoint families
%
\def\tenpoint{\def\rm{\fam0\tenrm}%
  \textfont0=\tenrm \scriptfont0=\sevenrm 
                      \scriptscriptfont0=\fiverm
  \textfont1=\teni  \scriptfont1=\seveni 
                      \scriptscriptfont1=\fivei
  \textfont2=\tensy \scriptfont2=\sevensy 
                      \scriptscriptfont2=\fivesy
  \textfont3=\tenex   \scriptfont3=\tenex 
                      \scriptscriptfont3=\tenex
  \textfont\itfam=\tenit  \def\it{\fam\itfam\tenit}%
  \textfont\slfam=\tensl  \def\sl{\fam\slfam\tensl}%
  \textfont\bffam=\tenbf  \scriptfont\bffam=\sevenbf
                            \scriptscriptfont\bffam=\fivebf
                            \def\bf{\fam\bffam\tenbf}%
  \normalbaselineskip=20 truept
  \setbox\strutbox=\hbox{\vrule height14pt depth6pt
width0pt}%
  \let\sc=\eightrm \normalbaselines\rm}
\def\eightpoint{\def\rm{\fam0\eightrm}%
  \textfont0=\eightrm \scriptfont0=\sixrm 
                      \scriptscriptfont0=\fiverm
  \textfont1=\eighti  \scriptfont1=\sixi
                      \scriptscriptfont1=\fivei
  \textfont2=\eightsy \scriptfont2=\sixsy
                      \scriptscriptfont2=\fivesy
  \textfont3=\tenex   \scriptfont3=\tenex
                      \scriptscriptfont3=\tenex
  \textfont\itfam=\eightit  \def\it{\fam\itfam\eightit}%
  \textfont\bffam=\eightbf  \def\bf{\fam\bffam\eightbf}%
  \normalbaselineskip=16 truept
  \setbox\strutbox=\hbox{\vrule height11pt depth5pt width0pt}}
\def\fourteenpoint{\def\rm{\fam0\fourteenrm}%
  \textfont0=\fourteenrm \scriptfont0=\tenrm 
                      \scriptscriptfont0=\eightrm
  \textfont1=\fourteeni  \scriptfont1=\teni 
                      \scriptscriptfont1=\eighti
  \textfont2=\fourteensy \scriptfont2=\tensy 
                      \scriptscriptfont2=\eightsy
  \textfont3=\tenex   \scriptfont3=\tenex 
                      \scriptscriptfont3=\tenex
  \textfont\itfam=\fourteenit  \def\it{\fam\itfam\fourteenit}%
  \textfont\bffam=\fourteenbf  \scriptfont\bffam=\tenbf
                             \scriptscriptfont\bffam=\eightbf
                             \def\bf{\fam\bffam\fourteenbf}%
  \normalbaselineskip=24 truept
  \setbox\strutbox=\hbox{\vrule height17pt depth7pt width0pt}%
  \let\sc=\tenrm \normalbaselines\rm}
\def\today{\number\day\ \ifcase\month\or
  January\or February\or March\or April\or May\or June\or
  July\or August\or September\or October\or November\or
December\fi
  \space \number\year}

%%%%%%%%%%%%%%%%%%%%%%%% Counter definitions %%%%%%%%%%%%%%%%%%%%%%%%
%
\newcount\secno      %section number
\newcount\subno      %number of subsection
\newcount\subsubno   %number of subsubsection
\newcount\appno      %appendix number
\newcount\tableno    %table number
\newcount\figureno   %figure number
%

%%%%%%%%%%%%%%%%%%%%%%%%%%%%% Baselineskip %%%%%%%%%%%%%%%%%%%%%%%%%%
%
\normalbaselineskip=20 truept
\baselineskip=20 truept

%%%%%%%%%%%%%%%%%%%% Specific formatting commands %%%%%%%%%%%%%%%%%%%
%
% Title of article
%
\def\title#1
   {\vglue1truein
   {\baselineskip=24 truept
    \pretolerance=10000
    \raggedright
    \noindent \fourteenpoint\bf #1\par}
    \vskip1truein minus36pt}
%

% Author names
% (The names of all the authors should be in the form initials then 
% surname. There should be no points after initials.)
%
\def\author#1
  {{\pretolerance=10000
    \raggedright
    \noindent {\large #1}\par}}

% Address of the authors
% (If authors are at differing addresses use one \address for each)
%
\def\address#1
   {\bigskip
    \noindent \rm #1\par}

% Short title (not more than fifty characters)
%
\def\shorttitle#1
   {\vfill
    \noindent \rm Short title: {\sl #1}\par
    \medskip}

% Physics Abstracts classification numbers
%
\def\pacs#1
   {\noindent \rm PACS number(s): #1\par
    \medskip}

% Journal article submitted to
%
\def\jnl#1
   {\noindent \rm Submitted to: {\sl #1}\par
    \medskip}

% Today's date
%
\def\date
   {\noindent Date: \today\par
    \medskip}

% Start of abstract
%
\def\beginabstract
   {\vfill\eject
    \noindent {\bf Abstract. }\rm}

% Keyword abstract - only required for J. Phys. G
%
\def\keyword#1
   {\bigskip
    \noindent {\bf Keyword abstract: }\rm#1}

% End of abstract
%
\def\endabstract
   {\par
    \vfill\eject}

% Contents page (only required for Reports on Progress in Physics)
%
% Heading for contents page
%

% Entry in list of contents (section headings)
%
\def\entry#1#2#3
   {\noindent
    \hangindent=20pt
    \hangafter=1
    \hbox to20pt{#1 \hss}#2\hfill #3\par}

% Subentry in list of contents (subsection heading).
% (Subsubsection headings do not appear in the contents list)
%
\def\subentry#1#2#3
   {\noindent
    \hangindent=40pt
    \hangafter=1
    \hskip20pt\hbox to20pt{#1 \hss}#2\hfill #3\par}
\def\checkforsub{\futurelet\nexttok\decide}
\def\ssf{\relax}
\def\decide{\if\nexttok\ssf\let\endspace=\nospace
                \else\let\endspace=\extraspace\fi\endspace}
\def\nospace{\nobreak\par\nobreak}
%
% Section heading (#1 is title of section, no number required)
%
\def\section#1{%
    \goodbreak
    \vskip50pt plus12pt minus12pt
    \nobreak
    \gdef\extraspace{\nobreak\bigskip\noindent\ignorespaces}%
    \noindent
    \subno=0 \subsubno=0 
    \global\advance\secno by 1
    \noindent {\bf \the\secno. #1}\par\checkforsub}

% Subsection heading (#1 is title of subsection, no number required)
%
\def\subsection#1{%
     \goodbreak
     \vskip24pt plus12pt minus6pt
     \nobreak
     \gdef\extraspace{\nobreak\medskip\noindent\ignorespaces}%
     \noindent
     \subsubno=0
     \global\advance\subno by 1
     \noindent {\sl \the\secno.\the\subno. #1\par}\checkforsub}

% Subsubsection heading (#1 is title of subsubsection, 
% no number is required)
%
\def\subsubsection#1{%
     \goodbreak
     \vskip20pt plus6pt minus6pt
     \nobreak\noindent
     \global\advance\subsubno by 1
     \noindent {\sl \the\secno.\the\subno.\the\subsubno. #1}\null. 
     \ignorespaces}

% Heading for an appendix, #1 is title of appendix, 
% no number or letter required
%
\def\appendix#1
   {\vskip0pt plus.1\vsize\penalty-250
    \vskip0pt plus-.1\vsize\vskip24pt plus12pt minus6pt
    \subno=0
    \global\advance\appno by 1
    \noindent {\bf Appendix \the\appno. #1\par}
    \bigskip
    \noindent}

% Heading for subsection within an appendix
%
\def\subappendix#1
   {\vskip-\lastskip
    \vskip36pt plus12pt minus12pt
    \bigbreak
    \global\advance\subno by 1
    \noindent {\sl \the\appno.\the\subno. #1\par}
    \nobreak
    \medskip
    \noindent}

% Heading for acknowledgments
%

%%%%%%%%%%%%%%%%%%%%%%%%%Macros for Tables%%%%%%%%%%%%%%%%%%%%%%%%%%%%

% Heading for start of tables section
%

% Table caption. #1 is caption, no number required
%
\def\tabcaption#1
   {\global\advance\tableno by 1
    \noindent {\bf Table \the\tableno.} \rm#1\par
    \bigskip}

% Definition of boldrule and medrule
%

% The halign (actually ialign) command for tables 
% (THIS MUST BE COPIED FOR EACH TABLE---WITHOUT THE PER CENT SIGN!)
% 
% \ialign{#\hfil&&\hglue 2pc plus2pc minus1pc#\hfil\cr

% A small negative skip for use in tables
%

% A macro for a footnote to a table
%

% Heading for list of figure captions
%

% Figure caption, #1 is caption, no number required
%
\def\figcaption#1
   {\global\advance\figureno by 1
    \noindent {\bf Figure \the\figureno.} \rm#1\par
    \bigskip}

% Heading for list of references
%
\def\references
     {\vfill\eject 
     {\noindent \bf References\par}
      \parindent=0pt
      \bigskip}

% Heading for list of numbered references
%

% Reference to a journal article in Harvard (alphabetical) system
%
\def\refjl#1#2#3#4
   {\hangindent=16pt
    \hangafter=1
    \rm #1
   {\frenchspacing\sl #2
    \bf #3}
    #4\par}

% Reference to a book or report in Harvard (alphabetical) system
%
\def\refbk#1#2#3
   {\hangindent=16pt
    \hangafter=1
    \rm #1
   {\frenchspacing\sl #2}
    #3\par}

% Reference to a journal article in numerical system
%
\def\numrefjl#1#2#3#4#5
   {\parindent=40pt
    \hang
    \noindent
    \rm {\hbox to 30truept{\hss #1\quad}}#2
   {\frenchspacing\sl #3\/
    \bf #4}
    #5\par\parindent=16pt}

% Reference to a book or report in numerical system
%
\def\numrefbk#1#2#3#4
   {\parindent=40pt
    \hang
    \noindent
    \rm {\hbox to 30truept{\hss #1\quad}}#2
   {\frenchspacing\sl #3\/}
    #4\par\parindent=16pt}

% Dash for use with repeated authors in reference lists
%

\def\ref#1{\noindent \hbox to 21pt{\hss 
#1\quad}\frenchspacing\ignorespaces}

% Fraction, alternative to \over
%
\def\frac#1#2{{#1 \over #2}}

% Renaming the dot under macro
%

% \d now used for differential d in mathematics
%

% \e gives roman e for exponential e in mathematics
%

% \i gives roman i for square root of minus one in maths mode
% and \ii used for dotless i in text mode

\chardef\ii="10

% Small (text size) fraction within displayed mathematics
%

% et al
%

% Redefinition of footnote macros to lose rule and remove indentation
%

\catcode`\@=11
\def\vfootnote#1{\insert\footins\bgroup
    \interlinepenalty=\interfootnotelinepenalty
    \splittopskip=\ht\strutbox % top baseline for broken footnotes
    \splitmaxdepth=\dp\strutbox \floatingpenalty=20000
    \leftskip=0pt \rightskip=0pt \spaceskip=0pt \xspaceskip=0pt
    \noindent\eightpoint\rm #1\ \ignorespaces\footstrut\futurelet\next\fo@t}

% Special macros for display equations
%
% \eq(#1) will give the equation number (#1) on the right
% instead of \eqno
%
\def\eq(#1){\hfill\llap{(#1)}}
\catcode`\@=12
%
% Macro for special accented characters
%
% Vectors with hats

% vectors with overbar

% roman characters with right pointing arrow

%
% Abbreviations for IOPP journals
%
\def\CQG{Classical Quantum Grav.}

        %1968-87
   %1988 and onwards
     %1968--1988
        %1989 and onwards

         %1968-89

           %1975--1988
     %1989 and onwards

         %1989 and onwards

%
% Other commonly quoted journals
%

%
% Miscellaneous definitions
%
% Bold nabla

%
% Bold dot for vector dot products
%

%
% Small space between lines in alignments or displayed maths

%
% Half line space for within tables or alignments

%
% Small negative space to close up lines above rules in tables

%
% greater than approximately signs
% \def\gap{\;\amex{\char'046}\;}           % use these if ams 
%  extension fonts available
% \def\sgap{\;\hbox{\samsx \char'046}\;}   % see above
\def\gap{\;\lower3pt\hbox{$\buildrel > \over \sim$}\;}
%
% less than approximately
%
% \def\lap{\;\amex{\char'056}\;}
% \def\slap{\;\hbox{\samsx \char'056}\;}
\def\lap{\;\lower3pt\hbox{$\buildrel < \over \sim$}\;}
% space between parts of short equations
\def\tqs{\hbox to 25pt{\hfil}}

   %order of

%
%\def\LaTeX{{\rm L\kern-.36em\raise.3ex\hbox{\eightrm a}\kern-.15em
%    T\kern-.1667em\lower.7ex\hbox{E}\kern-.125emX}}
%
% Better defn cribbed from `TeX for the Impatient':
%
\def\LaTeX{L\kern-.26em \raise.6ex\hbox{\fiverm A}%
   \kern-.15em\TeX}%
\def\AmSTeX{%
{$\cal{A}$}\kern-.1667em\lower.5ex\hbox{%
 $\cal{M}$}\kern-.125em{$\cal{S}$}-\TeX}

\def\D{\Delta} \def\gh{\hat g}
\def\gb{\bar g} \def\lh{\hat l}
\def\Db{\bar D} \def\Ivol{\int_\Sigma}
\def\Isurf{\int_{\partial \Sigma}}
 \def\sg{\sqrt{-g}}

\def\sf{\sqrt{1-kr^2}}

\title{Integral Constraints on Cosmological Perturbations and their Energy}
\author{Nathalie Deruelle$^{1,2}$, Joseph Katz$^3$, Jean-Philippe Uzan$^1$}
\address{$^1$ D\'epartement d'Astrophysique Relativiste et de Cosmologie,
UPR 176 du Centre National de la Recherche Scientifique,
Observatoire de Paris, 92195 Meudon, France}

\address{$^2$ DAMTP, University of Cambridge,
Silver Street, Cambridge, CB3 9EW, England}

\address{$^3$ Racah Institute of Physics, The Hebrew University,
Givat Ram, 91904, Jerusalem, Israel}

\shorttitle{}
\jnl{\CQG}
\date{}
\beginabstract
We show the relation between Traschen's integral equations and the
energy, and ``position of the centre of mass", of the matter
perturbations in a Robertson-Walker spacetime.  When perturbations are
``localised" we get a set of integral constraints that includes hers. We
illustrate them on a simple example.
\endabstract

\section{Introduction}

One ``puzzle" in the theory of cosmological
perturbations [1] is Traschen's ``integral constraints"
[2] (see also [3])~: besides the six standard
Robertson-Walker Killing vectors, she extracted from Einstein's
linearised equations four other vectors, that she called ``integral
constraint vectors". Each of those vectors yields an equation for the
matter perturbations, relating a volume to a surface integral. The
equations become constraints when perturbations are ``localised", for
which the surface integrals are zero.  Those constraints have been
widely used [4]--[7].  

A first question
we may ask is, {\it are there more than four such vectors ?} We will see
that the answer is ``yes", but that her vectors are particularly useful,
especially when perturbations are localised. Indeed the constraints she
obtains involve the matter variables only. However, other, simple,
constraints on the {\it geometry} exist as well, as we shall see in
Section 2.  

Second, several authors [2]--[4], [6] have interpreted Traschen's
equations as a generalisation of conservation laws for energy and
momentum in cosmology. Such quantities however are not straightforward
to define in general relativity. When Killing vector fields or an
asymptotic Killing vector fields exist, then of course we can write
integral quantities for the energy, momentum, angular momentum etc...
But Traschen's four ``integral constraint vectors" are not
Robertson-Walker Killing vectors. Thus, are we allowed to interpret the
conservation laws they imply as defining ``energy" and ``momentum"~?

A proper definition of conserved quantities such as energy, momentum
etc, involves the introduction of a background spacetime [8] and
hence depends a priori on the choice for the background, as well as on
the way points of the physical spacetime and of the background are
identified, i.e. on the mapping (see e.g. [9] and references
therein). Applying this formalism to perturbed Robertson-Walker
spacetimes (Section 3) we will first see how the choice of de Sitter
spacetime
as background is almost compulsory. Using its ten Killing vectors, we
will write ten Noether conservation laws, that is ten equations relating
volume to surface integrals. They will define, besides the known
momentum and angular momentum, an energy, $\delta E$, and a ``position
of the centre of mass", $\delta\vec Z$, of the perturbations of the
physical, perturbed Robertson-Walker spacetime. All will depend on the
constant $\bar R$ defining the de Sitter background and on the mapping.
We shall thus see that Traschen's integrals are {\it not} conserved
quantities. However, when the perturbations are localised,
Traschen's constraints are equivalent to $\delta E=0$ and $\delta \vec
Z=0$, independently of any mapping.  

The comparison between Traschen's integrals
and the conserved quantities is instructive in that it suggests to raise
to a special status a particular mapping in which Traschen's integral
constraint vectors become proportional to de Sitter Killing vectors (see
[10] for the mathematical origin of this property). This is done
in Section 4 where energy etc are expressed in that mapping, in a way
where all explicit reference to the background has disappeared.  

Finally,
in Section 5, we dwell on what is meant by ``localised" perturbations by
looking at the simple case of spherically symmetric perturbations. We
shall see that imposing the constraints amounts to imposing that not
only the matter perturbations, but also the metric perturbations, be
localised in space. Hence spacetime outside the perturbed region is
strictly Robertson-Walker and the constraints can, as already shown in
[1] on a Swiss cheese model, be interpreted as ``fitting
conditions" of the perturbed spacetime to a Robertson-Walker universe.
That also shows that the constraints hold only for perturbations which
are produced at some instant $t$ in a finite region of space and then
propagate in a up to then perfectly isotropic and homogeneous universe.
The origin of such perturbations cannot be described by Einstein's
equations~: they must arise from local processes like ``explosive"
events or phase transitions producing bubbles of true vacuum, cosmic
strings or other topological defects. And, indeed, it is in those
contexts that Traschen's constraints have been used [2], [5]--[7].

With this paper we hope to throw some light on
the meaning, and range of application, of integral constraints in
cosmology. We will also clarify the issue of defining energy, momentum
etc in spacetimes which are not asymptotically flat, in particular as
regards the role of background spacetimes and mappings in cosmology.

\section{Traschen's Vectors and Integral
Constraints on Cosmological Perturbations}

Traschen [4] has shown the existence of some vectors $V^\mu$ in
Robertson-Walker universes, which enter integral equations for arbitrary
perturbations. In this Section, we find them in a simple way and recall
what they are useful for. Let perturbed Robertson-Walker universes be
described in coordinates $x^\mu=(x^0\equiv t, x^k)$, ($\mu,\nu
...=0,1,2,3$~; $i,j,...=1,2,3$), such that the metric reads
$$ds^2=dt^2-a^2(t)(f_{ij} + h_{ij})dx^idx^j \eqno(1)$$
  $f_{ij}$ is the metric of a 3-sphere, plan
or hyperboloid depending on whether the index $k=(+1,0,-1)$ [11]
$$ f_{ij}=\delta_{ij}+k{\delta_{im}\delta_{jn}x^mx^n\over{1-kr^2}}
\qquad\hbox{with}\qquad r^2\equiv\delta_{ij}x^ix^j \eqno(2)
$$
the scale factor $a(t)$ is determined by Friedmann's equation and $h_{i
j }(x^\mu)$ is a small perturbation of $f_{ij}$.  We choose to work in a
synchronous gauge ($h_{00}=h_{i0}=0$) merely to simplify calculations (we
shall present gauge invariant calculations elsewhere). 

If $\delta T^\mu_\nu$ is the perturbation of the stress-energy tensor, the
linearised Einstein constraint equations read [12]
$$ \left\lbrace \matrix{\delta
G^0_0\equiv{1\over 2a^2}\left(\nabla_{m}\nabla_{n}\tilde h^{mn}+k\tilde
h\right)-{\dot a\over 2a}{\dot{\tilde h}} = \kappa \delta T ^0_0 \cr
\delta G^0_k\equiv{1\over2}\nabla_l {\dot {\tilde h}}^l_k=\kappa \delta
T^0_k \hfill \cr} \right. \eqno(3)
$$
$\kappa$ is Einstein's
constant, all indices are raised with the metric $f^{ij}$, $\nabla$
denotes the covariant derivative with respect to $f_{ij}$, a
dot denotes time derivative and we have introduced the
notation
$$ \tilde h_{ij}\equiv h_{ij}-f_{ij}h\quad\hbox{with}
     \quad h\equiv f^{ij}h_{ij}=-{1\over 2}\tilde h \eqno(4) $$
Let us now write equations (3) under an integral form
$$ {1\over\kappa}\Ivol\delta G^0_\mu\hat \zeta^\mu
d^3x=\Ivol\delta T^0_\mu \hat\zeta^\mu d^3x\qquad ,\qquad
{1\over\kappa}\Ivol\delta G^0_k\hat\zeta^k d^3x=\Ivol\delta T^0_k
\hat\zeta^k d^3x  \eqno(5)
$$
 $\zeta^\nu$ being an arbitrary vector
field~; a hat denotes multiplication by $\sqrt{-g}=
a^3\sqrt{f}=a^3/\sqrt{1-kr^2}$ (at zeroth order)~;  $\Sigma$ is a volume
in the hypersurface $t=Const$, and $d^3x \equiv dx^1dx^2dx^3$.  If we
perform the appropriate integrations by part to extract surface terms,
equations (3-5) read
$$ \eqalignno{
\Ivol \sqrt{-g}\left[ \delta T^0_\mu\zeta^\mu \right.
&+\left(\nabla^{(l}\zeta^{k)}
+Hf^{lk}\zeta^0\right){\dot{\tilde h}_{lk}\over2\kappa}  \cr
 &\left. -\left(\nabla^{lk}\zeta^0+kf^{lk}\zeta^0\right){\tilde h_{lk}\over 2
\kappa a^2}\right] d^3x
= \Isurf {\hat B}^l(\zeta) dS_l  &(6)  \cr }
$$
and
$$ \Ivol \sqrt{-g}\left( \delta
T^0_k\zeta^k +{1\over 2\kappa}\dot{\tilde h}^l_k
\nabla_l\zeta^k\right)d^3x={1\over \kappa}\Isurf\sqrt{-g} \dot{\tilde
h}^l_k\zeta^kdS_l  \eqno(7)
$$
$H$ is the Hubble parameter
$H\equiv\dot a/ a$, $\partial\Sigma$ is the boundary of the volume
$\Sigma$, $dS_k\equiv\epsilon_{klm}dx^{[l}dx^{m]}$, parentheses mean
symmetrisation, brackets antisymmetrisation, and
$$ B^l(\zeta)\equiv {1\over
2\kappa}\left[ {1\over a^2}\left(\zeta^0\nabla_k\tilde h^{kl}
-\tilde h^{ml}\nabla_m\zeta^0\right)+\zeta^k\dot{\tilde h}^l_k\right] \eqno(8)
$$
If $h_{ij}$ and $\partial_\rho h_{ij}$ vanish on the boundary $\partial\Sigma$,
the surface terms in (6-7) disappear, in which case equations (6-7)
become constraints (one for each vector $\zeta^\mu$) on the matter
perturbations $\delta T^0_\mu$.  

Equations (6-7) are identically
satisfied for all vector $\zeta^\mu$, if we take for the perturbations a
solution of the Einstein equations.  They simply relate a solution and
its boundary conditions.  Now, if one is looking for solutions
satisfying some particular boundary conditions (like localised
perturbations), then they {\it constrain} the set of solutions and can
give some of their properties.  Since $\zeta^\mu$ is a priori arbitrary,
there exists as many integral equations and constraints as independent
vector fields, that is an infinite number.  

However there are not so
many vector fields which can be considered as useful.  Indeed, for an
arbitrary $\zeta^\mu$, one needs the full metric $h_{kl}$ and hence one
must solve the full Einstein equations to compute the integrals.
Trashen's vectors $\zeta^\mu=V^\mu$ [4] are such
that
$$ \nabla^ { (l}V^{k)}+Hf^{lk}V^0=0\qquad ,\qquad
\nabla^{lk}V^0+kf^{lk}V^0=0  \eqno(9)
$$
so that the coefficients of
$\tilde h_{kl}$ and $\dot{\tilde h}_{lk}$ in (6) separately vanish.
Traschen's vectors therefore enable to decouple the perturbations of the
matter andthose of the geometry and give informations on the matter
perturbations (density, pressure...) alone, without having to solve the
full Einstein equations.  Indeed equation (6) then becomes
$$ \delta P_{Tr}(V) \equiv\int_\Sigma\delta T^0_\mu \hat
V^\mu \, d^3x = \int_{\partial\Sigma}\hat B^l(V) \, dS_l  \eqno(10)
$$
which are the ten Traschen's integral equations
[4].  As for Equation (7), it
becomes
$$ \int_\Sigma\sqrt{-g}\left(\delta
T^0_kV^k-{H\over2\kappa}\dot{\tilde
h}V^0\right)d^3x={1\over2\kappa}\int_{\partial\Sigma}\sqrt{-g}\dot{\tilde
h^l_k}V^kdS_l     \eqno(11)
$$

Traschen's vectors $V^\mu$ are ten linearly
independent, particular, solutions of equations (9) and any solution of
(9) is a linear combination of the $V^\mu$ with time dependent
coefficients. The explicit expressions of the $V^\mu$ are given in
Appendix 1. They split into two families. The first one contains the six
Robertson-Walker Killing vectors of spatial translations, $V^\mu=P^\mu$,
and spatial rotations, $V^\mu=R^\mu$ (see equations (A1-A2) for their
explicit expression). For those six vectors Traschen's equations (10) as
well as equations (11) are equivalent to
$$ \left\lbrace
\matrix{ \int_\Sigma\delta T^0_i d^3x={1\over2\kappa }
\int_{\partial\Sigma}\dot{\tilde h^l_i}dS_l\hfill &\;  {\rm for} \;  k=0 \cr
\int_\Sigma\delta
T^0_i{(\delta^{ir}x^s-\delta^{is}x^r)\over\sqrt{1-kr^2}}d^3x={1\over2\kappa}
\Isurf \dot{\tilde
h^l_i}{(\delta^{ir}x^s-\delta^{is}x^r)\over\sqrt{1-kr^2}} dS_l\hfil &\;  {\rm
for}\;k\neq 0 \cr } \right.  \eqno(12)
$$
The second family of
vectors contains four vectors, one $T^\mu$ and three $K^\mu$, that
Traschen called ``integral constraint vectors" or ICVs~:  see equations
(A6) and (A7) in Appendix~1 for their explicit expressions.  Note that
they are {\it not} Robertson-Walker conformal Killing vectors.  For those
four vectors Traschen's equations (10) read
$$ \delta
P_{Tr}(T)\equiv a^3 \Ivol \left(\delta\rho-H\delta T^0_lx^l\right)
d^3x=\Isurf\hat B^l(T) dS_l  \eqno(13)
$$
$$ \left\lbrace
\matrix{ \delta P^i_{Tr}(K)\equiv a^3 \Ivol\left[x^i\delta\rho
+H\delta
T^0_l\left(k\delta^{li}-x^lx^i\right)\right]{d^3x\over\sqrt{1-kr^2}}=\Isurf\hat
B^{li}(K) dS_l \;  & {\rm for} \;  k\neq0 \cr \delta
P^i_{Tr}(K)\equiv a^3\Ivol\left[x^i\delta\rho +H\delta T^0_l\left(
{1\over2}\delta^{li}r^2-x^lx^i\right)\right]d^3x=\Isurf\hat
B^{li}(K)dS_l \;  & {\rm for} \;  k=0 \cr }\right.\eqno(14)
$$
As for Equations (11) they become
$$ \Ivol \left(\delta T^0_lx^l+{1\over2\kappa}\dot{\tilde h}\right)
d^3x={1\over2\kappa}\Isurf\dot{\tilde h^l_k}x^kdS_l \eqno(15)
$$
$$ \left\lbrace \matrix{ \Ivol\left[\delta T^0_l
\left(k\delta^{li}-x^lx^i\right)-{1\over 2\kappa}
\dot{\tilde h}x^i\right]{d^3x\over\sqrt{1-kr^2}}={1\over2\kappa H}\Isurf
\dot{\tilde h^l_k}K^{ki}{dS_l\over\sqrt{1-kr^2}}
\; & \; {\rm for} \; k\neq0  \cr
\Ivol\left[\delta T^0_l\left( {1\over2}\delta^{li}r^2-x^lx^i\right)
   -{1\over2\kappa}\dot{\tilde h}
x^i\right]d^3x={1\over2\kappa H}\Isurf \dot{\tilde h^l_k} K^{ki}dS_l
\hfill & \; {\rm for} \; k=0  \cr }\right. \eqno(16)
$$
Since $T^\mu$ and $K^\mu$ are not Robertson-Walker Killing vectors the
interpretation of $\delta P_{Tr}(T)$ and $\delta P^
i_{Tr}(K)$ is not straightforward.  

Traschen considered perturbations that
are ``localised", for which the surface integrals vanish.
Equations (12-14) then become  constraints which read
$$ \int_\Sigma\delta T^0_i d^3x=0\qquad ,\qquad\int_\Sigma\delta
T^0_i{(\delta^{ir}x^s-\delta^{is}x^r)\over\sqrt{1-kr^2}}d^3x=0 \eqno(17)
$$
$$ \Ivol \left(\delta\rho-H\delta T^0_lx^l\right) d^3x=0 \eqno(18)
$$
$$ \left\lbrace \matrix{  \Ivol\left[x^i\delta\rho +H\delta
T^0_l\left(k\delta^{li}-x^lx^i\right)\right]{d^3x\over\sqrt{1-kr^2}}=0
\;  & \;  {\rm for} \;  k\neq0 \cr \Ivol\left[x^i\delta\rho +H\delta
T^0_l\left( {1\over2}\delta^{li}r^2-x^lx^i\right)\right]d^3x=0 \;  & \;
{\rm for} \;  k=0 \cr }\right.  \eqno(19)
$$
which are useful when studying localised (or ``causal") density
perturbations, especially when they are scalar that is such that $\delta
T^0_k=0$, which is the case in most practical applications
[2]--[4], [5]--[7].  

The constraints
(17-19) are the only ones which involve only the matter perturbations.
However when perturbations are localised Equations (15-16) also become
constraints
$$ \Ivol \left(\delta
T^0_lx^l+{1\over2\kappa}\dot{\tilde h}\right)
d^3x=0   \eqno(20)
$$
$$ \left\lbrace
\matrix{ \Ivol\left[\delta
T^0_l\left(k\delta^{li}-x^lx^i\right)-{1\over2\kappa}\dot{\tilde
h}x^i\right]{d^3x\over\sqrt{1-kr^2}}=0 \;  & \;  {\rm for} \;  k\neq0
\cr \Ivol\left[\delta T^0_l\left(
{1\over2}\delta^{li}r^2-x^lx^i\right)-{1\over2\kappa}\dot{\tilde
h}x^i\right]d^3x=0 \;  & \;  {\rm for} \;  k=0 \cr } \right.
\eqno(21)$$
This simple new constraints which involve only the
geometry when the perturbations are scalar could be useful in
numerical calculations.  We shall use them in a simple case in Section
5.

\section{Defining energy and
motion of the centre of mass in perturbed Robertson-Walker universes}

Several authors have interpreted equations (13-14) as
defining the energy and momentum of the perturbations of a
Robertson-Walker universe. However, to define properly energy,
momentum, angular momentum etc we shall introduce a background, as in
Katz [8] and Katz Bi${\check{\rm c}}$ak and Lynden-Bell
[9].  

Consider a spacetime $({\cal M}, g_{\mu\nu}(x^\lambda))$, a
background $(\bar{\cal M}, \bar g_{\mu\nu}(x^\lambda))$ and a mapping
between these two spacetimes, i.e. a way to identify points of ${\cal
M}$ and $\bar{\cal M}$.  

We take as lagrangian density for gravity
$$ {\hat {\cal L}}_{G}=\frac{1}{2\kappa}[{\hat g}^{\mu\nu}
(\Delta^\rho_{\mu\nu}\Delta^\sigma_{\rho\sigma} -
\Delta^\rho_{\mu\sigma}\Delta^\sigma_{\rho\nu}) -({\hat g}^{\mu\nu}
-\bar{\hat g}^{\mu\nu}){\bar R}_{\mu\nu}]
\eqno(22)$$
where we have
introduced the difference $\Delta^\lambda_{\mu\nu}$ between Christoffel
symbols in ${\cal M}$ and ${\bar{\cal M}}$ and where $\bar R_{\mu\nu}$
is the Ricci tensor of the background. We recall that a hat
denotes multiplication by $\sqrt{-g}$. $\hat {\cal L}_{G}$ vanishes when
$g_{\mu\nu} = {\bar g}_{\mu\nu}$, and is quadratic in the first order
derivatives of $g_{\mu\nu}$. It reduces to the familiar ``$\Gamma\Gamma
-\Gamma\Gamma$" form when the Riemann tensor of the background is zero
and when the coordinates are cartesian (such that
$\bar\Gamma^\lambda_{\mu\nu} = 0$). Since the ``$\Delta$" are tensors,
$\hat {\cal L}_{G}$ is a true scalar density.  

If we now perform a small
displacement $\Delta x^\mu = \zeta^\mu\Delta\lambda$, where
$\zeta^\mu$ is an arbitrary vector field and $\Delta\lambda$ an
infinitesimal parameter, and use the fact that $\hat {\cal L}_{G}$ is a
scalar density, we have that, with ${\rm L}_\zeta$ denoting the Lie
derivative,
$$ {\rm L}_\zeta \hat {\cal L}_G
-\partial_\mu(\hat {\cal L}_G \zeta^\mu)=0
\eqno(23)$$
Computing explicitely ${\rm L}_\zeta \hat {\cal L}_G$ from (22), it can
be shown (cf [9]) that there exists an identically conserved vector
$\hat I^\mu$ (that is such that $\partial_\mu \hat I^\mu\equiv 0$),
and hence an antisymmetric tensor $\hat J^{[\mu\nu]}$ such that
$$ \hat I^\mu = \partial_\nu \hat J^{[\mu\nu]}
\eqno(24)$$
The explicit expression for $\hat I^\mu$ is
$$ \hat I^\mu =\left[(\hat T^\mu_\nu - \bar{\hat T}^\mu_\nu)
+ \frac{1}{2\kappa}\hat l^{\rho\sigma}\bar R_{\rho\sigma}\delta^\mu_\nu
+ \hat t^\mu_\nu\right]\zeta^\nu +\hat \sigma^{\mu[\rho\sigma]}
  \partial_{[\rho}\zeta_{\sigma]} + {\hat Z}^\mu
(\zeta^\nu) \eqno(25) $$
with $\hat l^{\mu\nu}\equiv\hat g^{\mu\nu} - \bar {\hat g}^{\mu\nu}$.
(In equation (25) indices are moved with the background metric
$\bar g_{\mu\nu}$.) We can interpret the first term ($\hat T^\mu_\nu
- \bar {\hat T}^\mu_\nu$) as the energy-momentum tensor density of matter
with respect to the background.  The second term can be seen as a
coupling between the spacetime and the background.  The third one
reduces to the Einstein pseudo-tensor density when the background is flat
and the coordinates are cartesian.  The next term, quadratic in the
metric, is the helicity tensor density of the gravitational field with
respect to the background.  The last term is a function of the vectors
$\zeta^\mu$ which vanishes when those vectors are Killing vectors of
the background.  The explicit expressions for the various quantities
introduced, as well as that for $J^{[\mu\nu]}$ can be found in
[9] (see also Appendix 2).  

Let us stress that the equality $\hat
I^\mu(\{g^{\mu\nu},\bar g^{\mu\nu},\zeta^\nu\}) = \partial_\nu \hat
J^{[\mu\nu]}(\{g^{\mu\nu},\bar g^{\mu\nu}, \zeta^\nu\})$ and the
integral equation that can be deduced from it, are valid for all
$\{g^{\mu\nu},\bar g^{\mu\nu}, \zeta^\nu\}$.  We have written identities
which involve an arbitrary vector $\zeta^\mu$ (just as in Eq (5-7)),
two metrics and their derivatives.  Such identities are, in the
terminology of Bergman [13], {\it strong conservation laws}.
They reduce to the Noether conservation laws when the vectors $
\zeta^\mu$ become Killing vectors $\bar \xi^\mu$ of the background.

This means that, in order to obtain the maximum number of Noether
conservation laws, one must consider a background with maximal symmetry,
in which case ten integral equations (one for
each Killing vector $\zeta^\mu=\bar\xi^\mu$) can be written.
They are
$$ \eqalignno{
P(\bar\xi) \equiv \Ivol \hat I^\mu d\Sigma_\mu \equiv \Ivol
\left\{\left[(\hat T^\mu_\nu-\bar{\hat T}^\mu_\nu )  \right.\right.
& \left. + \frac{1}{2\kappa} \lh^{\rho\sigma}\bar R_{\rho\sigma}
\delta^\mu_\nu+\hat t^\mu_\nu\right]\bar\xi^\nu \cr
& \left. + \hat
\sigma^{\mu[\rho\sigma]}\partial_{[\rho} \bar\xi_{\sigma]}\right\}
d\Sigma_\mu \equiv \Isurf \hat J^{\mu\nu}d\Sigma_{\mu\nu}
&(26) \cr }
$$
where $d\Sigma_\mu$ is the volume element of a spacelike
hypersurface $\Sigma$, $d\Sigma_{\mu\nu}$ the surface element of its
boundary $\partial\Sigma$.  

We know two maximally symmetric spacetimes,
Minkowski and de Sitter spacetimes (we shall not consider here the
perhaps interesting anti-de Sitter possibility).  If
$\bar\xi^\mu=\bar T^\mu$ refers to the time translations in Minkowski
spacetime or the quasi-time translations of de Sitter spacetime, then the
quantity $P(\bar T)$ will be called energy.  When one uses the three
Killing vectors associated with the Lorentz rotations of Minkowski or
the quasi-Lorentz rotations of de Sitter spacetimes, $\bar \xi^\mu=\bar
K^\mu$, $P(\bar K)$ will be the ``position of the centre of mass"
[12].  The introduction of a maximally symmetric background thus
allows to define an energy etc, even if the physical spacetime does not
possess symmetries, globally or asymptotically.  The justification for
defining energy etc by (26) can be found in e.g. [9]. Minkowski
spacetime has been extensively used as background to study spacetimes
which are asymptotically flat (even if the role of the background is not
apparent, as is the case with pseudo-tensors when cartesian coordinates
are used from the start).  We want to define energy etc in cosmology,
and that will, as we shall see shortly, make us choose de Sitter rather
than Minkowski spacetime as background.  

We now apply the formalism
summarized above to a perturbed Robertson-Walker spacetime with metric
(1).  The maximally symmetric background will be chosen with the same
spatial topology as the physical perturbed Robertson-Walker spacetime
and the metric for the background will be written as
$$
d\bar s^2=\Psi(t)^2 dt^2 - \bar
a(t)^2f_{ij}dx^idx^j
\eqno(27)$$
Equation (27) contains a definition of
the mapping for each point of the $t=Const.$ hypersurface, up to an
isometry.  The function $\Psi(t)$ defines the mapping of the cosmic
times (and theexplicit expression for the scale factor $\bar a(t)$).
Those restrictions on the mapping render the choice of de Sitter rather
than Minkowski spacetime almost compulsory.  Indeed when the
Robertson-Walker sections are closed, Minkowski spacetime as background
is excluded, since there is no coordinate system in which it has closed
spatial sections.  When the Robertson-Walker sections are flat,
in order to have a one-to-one correspondance between $\Psi$ and $\bar a$,
Minkowski spacetime, which has $\bar a=1$ $\forall \Psi$ when $k=0$, must
again be excluded as background.  Hence, only when $k=-1$ is Minkowski
spacetime (in Milne coordinates) possible as background.  One may also
note that in the flat and open cases the physical spacetime is mapped on
only a patch of the de Sitter hyperboloid.  This is not a problem as we
need not fix the patch~:  the de Sitter Killing vectors corresponding
to the quasi-time translations and quasi-Lorentz rotations do not
apply the patch onto itself but displace it on the de Sitter
hyperboloid. 

The explicit expressions of the ten de Sitter Killing
vectors when the metric is written under the form (27) are given in
Appendix 1 (equations (A1-5)).  They satisfy equations very similar to
the equations (9) satisfied by Traschen's ICVs, to
wit~:
$$ \nabla^{(l}\bar\xi^{k)}+{\dot{\bar a}\over\bar
a}f^{lk}\bar\xi^0=0\qquad,\qquad\nabla^{lk}\bar\xi^0+kf^{lk }
\bar\xi^0=0 \eqno(28)
$$

The zeroth order conservation quantities
$P_{RW}(\bar \xi)$ have been defined and studied by Katz Bi$\check {\rm
c}$ak and Lynden-Bell [9].  Here we focus on their perturbations at
first order.  A fairly long but straightforward calculation brings
equation (26) to the form~:  $P(\bar\xi)=P_{RW}(\bar\xi) +\delta
P(\bar\xi)$, where $\delta P(\bar\xi)$ is the sum of equation (6), with
$\zeta^\mu$ a de Sitter Killing vector satisfying equation (28), and
a surface term
$$  \delta P(\bar\xi) \equiv\Ivol
\sqrt{-g}\left(\delta T^0_\mu\bar\xi^\mu +{1\over2} \beta{\dot{\tilde
h}}\bar\xi^0\right)d^3x + \Isurf\hat M^l(\bar\xi)dS_l= \Isurf (\hat
B^l+\hat M^l)(\bar\xi)dS_l \eqno(29)
$$
 where
$$ M^l(\bar\xi)\equiv \frac{h}{2\kappa}\left\{\left[-2H
+\frac{\bar H}{2\Psi} \left(\frac{{\bar
a}^2}{a^2}+3\Psi^2\right)\right]{\bar \xi}^l +\frac{1}{2}\left(\Psi^2 -
\frac{{\bar a}^2}{a^2}\right) \frac{f^{kl}}{{\bar a}^2} \nabla_k {\bar
\xi}^0 \right\} \eqno(30)
$$
$B^l$ is given by equation (8) and we have introduced the notation
$$
\kappa\beta \equiv \frac{\dot a}{a} - \frac{\dot {\bar
a}}{\bar a} \eqno(31)
$$
as well as the Hubble parameter of the
background $\bar H\equiv\dot{\bar a}/\Psi\bar a$. 

Using the explicit
expressions of the De Sitter/Robertson-Walker Killing vectors
corresponding to spatial translations, $\bar\xi^\mu=P^\mu$ (see equation
(A1)), the total linear momentum of the perturbations is thus defined
as
$$ \delta P_i(P)\equiv a^3
\int_\Sigma d^3x\delta T^0_i+\int_{\partial\Sigma}\hat
M^l_i(P)dS_l=\int_{\partial\Sigma}(\hat B^l_i+\hat
M^l_i)(P)dS_l \eqno(32)
$$
with
$$ \hat
M^l_i(P)={a^3h\over2\kappa}\left[-2H+{\bar H\over2\Psi}\left({\bar
a^2\over a^2}+3\Psi^2\right)\right]\delta^l_i\quad ,\quad \hat
B^l_i(P)={a^3\over 2\kappa}\dot{\tilde h^l_i} \eqno(33)
$$
and a similar
expression for their total angular momentum corresponding
to $\bar\xi^\mu=R^\mu$ as given by equation (A2).  One sees that the
total linear (and angular) momentum is the sum of a background and
mapping independent volume integral plus a surface term which does depend
on the background and the mapping.  

When perturbations are
localised equation (32) becomes a constraint which is Traschen's
constraint (17). When it comes now to the de Sitter Killing vectors
corresponding to quasi-time translations ($\bar\xi^\mu=\bar T^\mu$) and
quasi-Lorentz rotations ($\bar\xi^\mu=\bar K^\mu$), not only the
definitions, as written in (29-30), of the corresponding energy and motion
 of the centre of mass of the perturbations, but also the constraints
which follow when the perturbations are localised, seem to depend on the
background and the mapping. Now, that the {\it definition} of conserved
quantities be dependent on conventions for the choice of background or
mapping is not a problem. On the other hand {\it constraints},
which contain measurable information (for example they imply a drastic
reduction of the cosmic microwave background
anisotropies [2], [5]) cannot be mapping dependent. To show
explicitely that indeed thedefinition of energy and motion of the centre
of mass depends on the mapping and the background (just as the total
linear and angular momentum), but that the constraints do not,
we rearrange equations (29-30) using the explicit expressions (A3-A5) for
$\bar T^\mu$ and $\bar K^\mu$ as well as the relations (15-16) to
eliminate $\dot{\tilde h}$ in the volume integral of (28). We obtain
$$  \delta E = \frac{1}{\Psi}\delta P_{Tr}(T) +\Isurf
\left[{ \hat M}^l(\bar T) +\frac{\sg}{2\kappa\Psi} \left(1-\frac{{\dot
a}{\bar a}}{a{\dot{\bar a}}}\right) {\dot {\tilde h}}^l_k \bar T^k
\right]dS_l \eqno(34)
$$
$$  \left\lbrace \matrix{
\delta Z^i =\frac{1}{\Psi} \delta P^i_{Tr}(K) + \Isurf \left[{\hat
M}^{li}(\bar K) +\frac{\sg}{2\kappa\Psi} \left(1-\frac{{\dot a}{\bar
a}}{a{\dot{\bar a}}}\right){\dot {\tilde h}}^l_k \bar K^{ik} \right]dS_l
\hfill & {\rm for} \;  k\neq0 \cr \delta Z^i = \frac{1}{\Psi}\delta
P^i_{Tr}(K) - \frac{ 1 }{2{\bar H}{\bar a}^2}\delta P^i(P) +\Isurf
\left[{\hat M}^{il}(\bar K) +\frac{\sg}{2\kappa\Psi} \left(1-\frac{{\dot
a}{\bar a}}{a{\dot{\bar a}}}\right) {\dot {\tilde h}}^l_k \bar
K^{ik}\right]dS_l & {\rm for} \;  k=0 \cr }\right.
\eqno(35)$$

We have introduced the short-hand notation $\delta E\equiv\delta P(\bar
T)$ and $\delta Z^i\equiv\delta P^i(\bar K)$, and the background and
mapping independent $\delta P_{Tr}$ are given by equations (13-14).
Hence, the energy and motion of the centre of mass of the perturbations
are the sum of volume integrals which are, up to the overall function of
time $\Psi$, background and mapping independent, plus surface terms which
do depend on the background and the mapping.  We thus see the announced
relationship between the energy and motion of the centre of mass of the
perturbations and Traschen's integrals (10) (13-14).  Turning to
localised perturbations for which all surface integrals vanish, we
finally see on the form (34-35) for the conserved quantities
that the resulting constraints are background and mapping independent,
and are Traschen's constraints (18-19).

\section{Mapping the cosmic times}

 The conserved quantities defined
in the previous section are background and mapping dependent. We show
in this section that there is a mapping of the cosmic times of
particular significance. To see that, we shall use the relationship
found in [9] between Traschen's ICVs $V^\mu$ and de Sitter Killing
vectors $\bar\xi^\mu$. The four ICVs which are not de Sitter Killing
vectors are given by equations (A6-A7); as for the four de
Sitter Killing vectors corresponding to quasi-time translations and
quasi-Lorentz rotations they are given by equations (A3-A5), so that we
have
$$ T^0=\Psi \bar T^0\qquad ,\qquad T^k={ H\over\bar
H}\bar T^k \eqno(36)
$$
$$\left\lbrace \matrix{
K^0=\Psi \bar K^0\qquad ,\qquad K^k={ H\over\bar H}\bar K^k \;  & \;
{\rm for} \;k\neq 0 \cr K^0=\Psi\bar K^0\qquad ,\qquad K^k={ H\over\bar
H}\left(\bar K^k+{1\over2\bar H\bar a^2}P^k\right) \;  & \;  {\rm for} \;
k=0 \cr }\right. \eqno(37)
$$
(As for the remaining six
Traschen and de Sitter Killing vectors, they are identical and correspond
to the six Robertson-Walker Killing vectors $P^\mu$ and $R^\mu$.  We
also note that Traschen's ICVs become combinations of full-fledged
Killing vectors when the Robertson-Walker spacetime becomes a de Sitter
spacetime [5], [10].) 

Now, as emphasised in [9], in the
particular mapping
$$ a=\bar a\qquad \Longrightarrow\qquad
\Psi=H/\bar H\quad\hbox{and}\quad \beta=0
\eqno(38)$$
Traschen's ICVs
become (for $k=\pm 1$) strictly proportional to the de Sitter Killing
vectors~:  $V^\mu=\Psi\bar\xi^\mu$, where the function $\Psi$ is
completely determined once the Robertson-Walker scale factor $a(t)$ is
known.  For example, in the case of flat spatial sections,
$\Psi=2H\sqrt{3 /\bar R}$, where $\bar R$ is the scalar curvature of the
de Sitter background.

This property suggests to raise the mapping (38)
to a special status. Moreover the surface terms then acquire a
particularly simple form. If, finally, one normalises $\bar R$ to 12, for
which $\bar H=1$ when $k=0$, then all explicit reference to the de Sitter
background disappears from the definitions (34-35).  For example, the
energy of the perturbations of a flat Roberston-Walker spacetime becomes
$$  \delta E\equiv {a^3\over H}\Ivol(\delta \rho-H\delta
T^0_lx^l)d^3x +{a^3\over H}{H^2-1\over 4\kappa}\Isurf hx^ldS_l
\eqno(39)$$

\section{Integral constraints and ``localised perturbations"}

Ellis and Jaklitsch [1] have given an interpretation of Traschen's
Integral Constraints in terms of ``fitting conditions", using as an
example the ``Swiss cheese" model. We shall do the same for another
simple case, that of spherical perturbations. This will clarify further
what is meant by ``localised" perturbations and examplify the use of our
constraints (20-21). 

Consider a spherical symmetric perturbation of a
spatially flat dust universe.  Spherical perturbations are scalar.  The
integral equations (13) (15) for $\delta\rho$ and $\dot h$ reduce to
$$ \int_0^R \dot h r^2 dr= G(R) \quad ,\quad \int_0^R\delta
    \rho r^2dr= F(R) \eqno(40) $$
where $R$ is the radius of the sphere
on which the integration is performed and where $G$ and $F$ are some
fonctions of the metric perturbations and their derivatives.  Imposing
that perturbations be localised has meant, in the context of this
paper, that the surface terms in (40) be zero for all surfaces outside a
sphere of radius one, say.  The integral equations then become
constraints,
$$ \int_0^R \dot h r^2 dr= \int_0^R\delta
\rho\, r^2dr=0\quad\forall\quad R>1 \eqno(41)
$$
which imply
$$ \delta\rho=\dot h=0\quad\forall\quad
r>1\qquad\Rightarrow\qquad h=h(r)\quad {\rm for}\quad r>1 \eqno(42)
$$

However, imposing (41) means more than just (42).  To see that let's go
back to Einstein's equations.  Their solution is known.  It is the
linearisation of a Tolman-Bondi solution [14].  It depends on two
arbitrary functions $t_0(r)$, the delayed Big-Bang, and $\epsilon(r)$,
the local curvature.  A flat Robertson-Walker universe corresponds
to $t_0=\epsilon=0$.  In the case where $t_0(r)=0$ and $\epsilon(r)<<1$,
the metric reads, with $a(t)\equiv(\frac{3}{2}t)^{2/3}$
$$ ds^2=dt^2 -a^2(t)\left\{\left[1+r^2\epsilon(r)
-\frac{2}{3}a(t) (\epsilon + r\epsilon')\right]dr^2 + \left[1- 
\frac{2}{3}a(t)\epsilon\right]r^2d\Omega\right\} \eqno(43)
$$
 and we have
$$ \left \lbrace\matrix{ \delta \rho(r,t) &=&
\frac{1}{\kappa a(t)^2} \Xi(r) \hfill\cr h(r,t) &=& \frac{2}{3}a(t)\Xi(r) -
r^2\epsilon(r) \cr {\dot h}(r,t)
 &=& \frac{2}{3}{\dot a}(t)\Xi(r) \hfill \cr }\right. \eqno(44)
$$
where we have introduced the function
$\Xi(r)=\frac{1}{r^2}(r^3\epsilon)'$.  

Therefore, the conditions (42)
only amount to imposing that $\Xi(r)=0$ for $r>1$ or, equivalently, that
the perturbation $\epsilon(r)$ be of the form
$$ \epsilon(r)=\epsilon(1)/r^3\quad\forall\quad
r>1\qquad\Rightarrow\qquad h=-\epsilon(1)/r\quad\forall\quad
r>1 \eqno(45)
$$
where $\epsilon(1)$ is a constant, whereas the
stronger constraints (41) add the extra condition
$$ \epsilon(1)=0\qquad\Rightarrow\qquad h=0\quad\forall\quad r>1\eqno(46)$$

 We therefore see on this simple example that
``localised" perturbations, that is perturbations such that the surface
terms vanish outside a certain region, are not simply perturbations
for which $\delta\rho=0$, but perturbations for which $\delta\rho=0$ and
$h_{ij}=0$ outside a certain region.  Outside that region, spacetime is
strictly Robertson-Walker. Hence, the constraints hold only for
perturbations that arise from local processes like``explosive" events, or
phase transitions producing bubbles of true vacuum, cosmic strings
or other topological defects [2], [5--7].We
can thus interpret the constraints in the following way~:  if
spacetime is strictly Robertson-Walker outside a certain region, then the
metric can be chosen so that $h_{ij}=0$ outside that region, and
Einstein's equations then tell us that the conserved quantities of
the perturbations inside that region are all zero.  Moreover, since the
``background" scale factor, $a(t)$, is the same as that of the outside
Robertson-Walker universe, the constraints can also be interpreted as
``fitting" conditions [1].

\appendix
{The ten de Sitter Killing vectors and Traschen's ICVs}

We write the de Sitter metric as~:
$$ d\bar s^2=\Psi^2(t)dt^2-\bar a^2(t)f_{ij}dx^idx^j $$
where $\Psi$ is an arbitrary function of time and
where $x^i$ are Weinberg's [11] coordinates, so that
$$ f_{ij}=\delta_{ij}+k{\delta_{im}\delta_{jn}x^mx^n\over1-kr^2}\quad{\rm
    with}\quad r^2=\delta_{ij}x^ix^j$$
where $k=+1,0,-1$ depending on
whether the spatial sections are closed, flat, or hyperbolic.  We have
that $\sqrt{f}=1/\sf$.  Let us also introduce the quantities
$$\bar H\equiv{1\over\Psi}{\dot{\bar a}\over\bar a}\qquad ,\qquad
      \tau\equiv{\Psi\over\dot{\bar a}}$$

Ten independent Killing vectors
describe three spatial translations, three spatial rotations, one
quasi-time translation and three quasi-Lorentz rotations.  Their
explicit expression is [9] 

(a) spatial translations~: $\bar\xi^\mu=P^\mu$
$$ P^0=0\qquad ,\qquad
     P^k=\delta^k_r\sqrt{1-kr^2}\eqno(A1)$$

(b) spatial rotations~:  $\bar\xi^\mu=R^\mu$
$$R^0=0\qquad ,\qquad R^k=\delta^{kr } x^s-\delta^{ks}x^r\eqno(A2)$$

(c) quasi-time translations~:  ${\bar \xi}^\mu=\bar T^\mu$
$$\bar T^0={1\over\Psi}\sqrt{1-kr^2}\qquad ,\qquad\bar T^k= -\bar
   Hx^k\sqrt{1-kr^2}\eqno(A3)$$

(d) quasi-Lorentz rotations~:
    $\bar\xi^\mu=\bar K^\mu$
$$\bar K^0={1\over \Psi}x^r\quad ,\quad\bar K^k=\bar
   H(k\delta^{kr}-x^kx^r)\qquad {\rm if}\quad k=\pm1\eqno(A4)$$
$$\bar K^0={1\over\Psi}x^r\quad ,\quad\bar K^k=\bar
   H\left[{1\over2}\delta^{kr}(r^2-\tau^2)-x^kx^r\right]\qquad {\rm
   if}\quad k=0\eqno(A5)$$

In analogy with special relativistic
definitions [12], the conserved quantity corresponding to spatial
translations is momentum, angular momentum corresponds to spatial
rotations, energy to quasi-time translations and position of the centre
of mass to quasi-Lorentz
rotations.  

We write Roberston-Walker metrics as
$$ds^2=dt^2- a^2(t)f_{ij}dx^idx^j$$
We also introduce
$$H\equiv{\dot a\over a}$$
With these coordinates we also have that $\sg = a^3\sqrt{f} =a^3/\sf$.

In that coordinate system, six of the ten Traschen vectors $V^\mu$ are
nothing but the previous Robertson-Walker/de Sitter Killing vectors
corresponding to spatial translations and rotations.  The extra four,
the ``integral constraint vectors" $T^\mu$ and $K^\mu$, read
$$T^0=\sqrt{1-kr^2}\quad ,\qquad T^ k =-Hx^k\sqrt{1-kr^2}\eqno(A6)$$
$$ K^0=x^r\quad ,\qquad K^k=H (k\delta^{kr}-x^kx^r)\qquad {\rm for}\quad
         k=\pm1\eqno(A7) $$
$$K^0 = x^r\quad ,\qquad K^k=
    H\left({1\over2}\delta^{kr}r^2-x^kx^r\right)\qquad {\rm for}\qquad
    k=0\eqno(A8) $$

They are related to the de Sitter Killing vectors
corresponding to quasi-time translations, quasi-Lorentz rotations, and,
in the flat case, spatial translations, by [9]
$$T^0=\Psi\bar T^0\qquad ,\qquad T^k={H\over\bar H}\bar
       T^k\qquad\forall\quad k\eqno(A9)$$
$$ K ^0=\Psi\bar K^0\qquad ,\qquad K^k={H\over\bar H}\bar
    K^k\qquad ,\qquad {\rm for}\quad k=\pm1\eqno(A10)$$
$$K^0 = \Psi\bar
K^0\qquad ,\qquad K^k= \frac{H}{\bar H}\left(\bar K^k+{1\over2\bar H\bar
a^2}P^k\right)\qquad {\rm for}\qquad k=0\eqno(A11)$$

All vectors
${\tilde V}^\mu=F(t)V^\mu$, with $F(t)$ an arbitrary function of time,
are solutions of Traschen's equations~(9).  If one chooses
$F(t)\equiv\Psi^{-1}$, and the function $\Psi$ such that $H=\Psi{\bar
H}$, then the vectors ${\tilde V}^\mu$ become a combination of the de
Sitter Killing vectors.

\appendix{}

One can extract frome Einstein's equations a conserved
current ${\hat I}^\mu$ and an anti-symmetric tensor ${\hat J}^{\mu\nu}$
such that
$${\hat I}^\mu = \partial_\nu {\hat J}^{[\mu\nu]} \qquad
\Leftrightarrow \qquad \partial_\mu {\hat I}^\mu = 0\eqno(B1)$$
We give here the expressions of these two quantities [see ref.[9]].

First introduce
$$\lh^{\mu\nu} \equiv \gh^{\mu\nu} -{\bar {\gh} }
^{\mu\nu} \qquad {\rm and} \qquad \D^\lambda_{\mu\nu} \equiv
\Gamma^\lambda_{\mu\nu} - {\bar \Gamma}^\lambda_{\mu\nu}\eqno(B2)$$
where $\Gamma$ and ${\bar \Gamma}$ are the Christoffel symbols of the
spacetime and of the background, ${\bar D}_\mu$ and
$D_\mu$ the two covariant derivatives and for a given quantity $A$,
${\hat A}$ denotes $\sg A$ and ${\bar A}$ the value of $A$ on the
background.  Notice that ${\bar {\hat A}} = \sqrt{-{\bar g}}{\bar A}
\not= {\hat {\bar A}} = \sqrt{-g}{\bar A}$.

\noindent {\bf expression for ${\hat J}^{\mu\nu}$}

For any vector $\zeta^\mu$ we have,
$$\kappa {\hat J}^{\mu\nu} = \lh^{[\mu\rho}\Db_\rho { \zeta}^{\nu]}
+\gh^{[\mu\rho}\D^{\nu]}_{\rho\lambda} { \zeta}^{\lambda}
+{\zeta}^{[\mu}\gh^{\nu]\rho} \D^ \sigma_{\rho\sigma} -{\zeta}^{[\mu}
\D^{\nu]}_{\rho\sigma}\gh^{\rho\sigma}\eqno(B3)$$

\noindent {\bf expression for ${\hat I}^\mu$ }

$${\hat I}^\mu = \left[\left({\hat
T}^\mu_\nu - {\bar {\hat T}^\mu_\nu}\right) + \frac{1}{2\kappa}{\hat
l}^{\rho\sigma}{\bar R}_{\rho\sigma}\delta^\mu_\nu +{\hat
t}^\mu_\nu\right]\zeta^\nu +{\hat
\sigma}^{\mu[\rho\sigma]}\partial_{[\rho}{ \zeta}_{\sigma]} + {\hat
Z}^\mu(\zeta^\nu)\eqno(B4)$$

\noindent ${\hat T}^\mu_\nu$ and ${\bar
{\hat T}^\mu_\nu}$ are the two energy-momentum tensors.
$$\eqalignno{
 2\kappa
  {\hat t}^\mu_\nu = \gh^{\rho\sigma}
  \left(\D^\lambda_{\rho\lambda}\D^\mu_{\sigma\nu} +
  \D^\mu_{\rho\sigma}\D^\lambda_{\lambda\nu} -
  2\D^\mu_{\rho\lambda}\D^\lambda_{\sigma\nu}\right)
&+ \gh^{\mu\rho}\left(\D^\sigma_{ \lambda\sigma}\D^\lambda_{\rho\nu} -
   \D^\sigma_{\sigma\rho}\D^\lambda_{\lambda\nu}\right)\qquad\qquad \cr
&-\gh^{\rho\sigma}
   \left(\D^\eta_{\rho\sigma}\D^\lambda_{\lambda\eta} -
  \D^\eta_{\rho\lambda}
  \D^\lambda_{\eta\sigma}\right)\delta^\mu_\nu &(B5) \cr }   $$

This term reduces
to the Einstein pseudo-tensor density when the background is
Minkowski spacetime in cartesian coordinates.
$$2\kappa {\hat
\sigma}^{\mu[\rho\sigma]} =\left(\lh^{\mu[\rho} \gb^{\sigma]\lambda} -
\gb^{\mu[\rho}\lh^{\sigma]\lambda}\right) \D^\nu_{\lambda\nu} -2
\lh^{\lambda[\rho}\gb^{\sigma]\nu} \D^\mu_{\lambda\nu}\eqno(B6)$$
$$ \eqalignno{
4\kappa {\hat Z}^\mu (\zeta^\nu) = \left(Z^\mu_\rho \gh^{\rho\sigma} +
  \gh^{\mu\rho}Z^\sigma_\rho
  -\gh^{\mu\sigma}Z\right)\D^\lambda_{\sigma\lambda}
 &+\left(\gh^{\rho\sigma}Z - 2 \gh^{\rho\lambda}Z^\sigma_\lambda\right)
  \D^\mu_{\rho\sigma} +\gh^{\mu\lambda}\partial_\lambda Z   \cr
&+\gh^{\rho\sigma}\left(\Db^\mu Z_{\rho\sigma} - 2\Db_\rho
    Z^\mu_\sigma\right) &(B7) \cr }$$
with
$$Z_{\rho\sigma} = {\rm L}_\zeta
\gb_{\rho\sigma} = \Db_{(\rho} \zeta_{\sigma)}\;\;  {\rm and} \;\;  Z
= Z_{\rho\sigma} \gb^{\rho\sigma}\eqno(B8)$$
When $\zeta^\mu$ is a
Killing vector of the background, $Z_{\rho\sigma}=0$ and thus $Z^\mu=0$.

\references
\numrefjl{[1]}{Ellis G F R
and Jaklitsch M J 1989}{Astrophysical Journal}{346}{601}
\numrefjl{[2]}{Traschen J 1984}{Phys. Rev. D}{29}{1563}
\numrefjl{[3]}{Abbott L F,
Traschen J, Xu Rui-Ming 1988}{Nucl. Phys.}{B296}{710}
\numrefjl{[4]}{Traschen J 1985}{Phys. Rev. D}{31}{283}
\numrefjl{[5]}{Traschen J and Eardley D 1986}{Phys. Rev. D}{34}{1665}
\numrefjl{[6]}{
Veeraraghavan S and Stebbins A 1990}{Astrophysical Journal}{365}{37-65}
\numrefjl{[7]}{Traschen J,
Turok J and Brandenberger R 1986}{Phys. Rev. D}{34}{919}
\numrefbk{[8]} {Katz J 1996 in}{Gravitational Dynamics}{edited by Lahav,
Terlevitch and Terlevitch (Cambridge University Press, p.  193)}
\numrefjl{[9]}
{Katz J, Bi${\check {\rm c}}$ak J, Lynden-Bell D 1996}{``Relativistic
conservation laws and integral constraints in cosmology}{preprint}
\numrefjl{[10]}{Todd K P 1988}{Class. and Quant.  Grav.}{20}{1297}
\numrefbk{[11]}{Weinberg S 1972 in}{Gravitation and Cosmology}{
(Wiley: New-York, Chap.  13)}
\numrefbk{[12]}{Landau L D
and Lifchitz E M 1972 in}{Th\'eorie Classique des Champs}{Mir}
\numrefjl{[13]}{Bergmann P G 1949}{Phys.  Rev.}{75}{680}
\numrefjl{[14]}{Tolman R C 1934}{Proc. Nat. Acad. Sci. USA}{20}{169}

\bye